\documentclass[aps,prl,reprint,amsmath,amssymb,showpacs]{revtex4-1}
\usepackage{graphicx}% Include figure files

\begin{document}
%opening
\title{Experimental Control of Transport and Current Reversals in a Deterministic Optical Rocking Ratchet}

\author{Alejandro \surname{V. Arzola}}
\email[]{avarzola@gmail.com}
\affiliation{Instituto de F\'isica, Universidad Nacional Aut\'onoma de M\'exico, Apdo. Postal 20-364, 01000 M\'exico, D.F., M\'exico}
\author{Karen \surname{Volke-Sep\'ulveda}}
\email[]{karen@fisica.unam.mx}
\affiliation{Instituto de F\'isica, Universidad Nacional Aut\'onoma de M\'exico, Apdo. Postal 20-364, 01000 M\'exico, D.F., M\'exico}
\author{Jos\'e \surname{L. Mateos}}
\affiliation{Instituto de F\'isica, Universidad Nacional Aut\'onoma de M\'exico, Apdo. Postal 20-364, 01000 M\'exico, D.F., M\'exico}

\begin{abstract}
We present an experimental demonstration of a deterministic optical rocking ratchet. A periodic and asymmetric light pattern is created to interact with dielectric microparticles in water, giving rise to a ratchet potential. The sample is moved with respect to the pattern with an unbiased time-periodic rocking function, which tilts the potential in alternating opposite directions. We obtain a current of particles whose direction can be controlled in real time and show that particles of different sizes may experience opposite currents. Moreover, we observed current reversals as a function of the magnitude and period of the rocking force.
\end{abstract}
%\keywords{Ratchets, optical trapping, optical sorting.}
\pacs{87.80.Cc, 05.60.Cd, 05.45.-a, 82.70.Dd}
\maketitle
The study of transport induced by symmetry breaking under unbiased forces has flourished as one of the most active and diverse fields in recent times. It includes the study of the so called Brownian motors and ratchets, initially motivated by the transport of molecular motors in the biological realm, but soon extended to many other domains in classical and quantum physics: single-particle transport, cold atoms in optical lattices, superconducting devices, granular flows, colloidal sorting, to name but a few \cite{Hanggirev09}. Among the many kinds of ratchets, an import class refers to classical deterministic ratchets in which the dynamics does not have any randomness or stochastic elements \cite{JLMat2000}. The paradigmatic model is a classical particle in a periodic asymmetric (ratchet) potential, acted upon by an additional external time-dependent force of zero average. If this external force is additive, we are considering a rocking ratchet. 

There have been some experiments using optical lattices to trap colloidal brownian particles, in order to obtain a systematic transport in the presence of unbiased forces (ratchet effect) \cite{Faucheux95, SHyuk, Lopez2008}. In these cases, the amplitude of the periodic potential is modulated in time, corresponding to the so-called flashing or pulsating ratchet. On the other hand, the ratchet effect has been obtained for symmetric optical lattices with an asymmetric time-dependent rocking force, and also for asymmetric optical potential with a pulsating activation, but in the quantum domain and in the inertial regime \cite{RGommers2006, Salger2009}. However, there are interesting predicted phenomena for rocking ratchets in the classical deterministic and overdamped regime \cite{Reimann98, aldana2006, Zarlenga09}, which have not been observed so far. In this Letter we will describe an experimental model of such a ratchet and show that we are able to obtain non-trivial particle transport, whose direction can be controlled in real time as a function of different experimental parameters. Furthermore, we present the first experimental verification of the current reversals in this regime, predicted since 1998 \cite{Reimann98}.

\begin{figure}
\includegraphics[angle=0,width=3in]{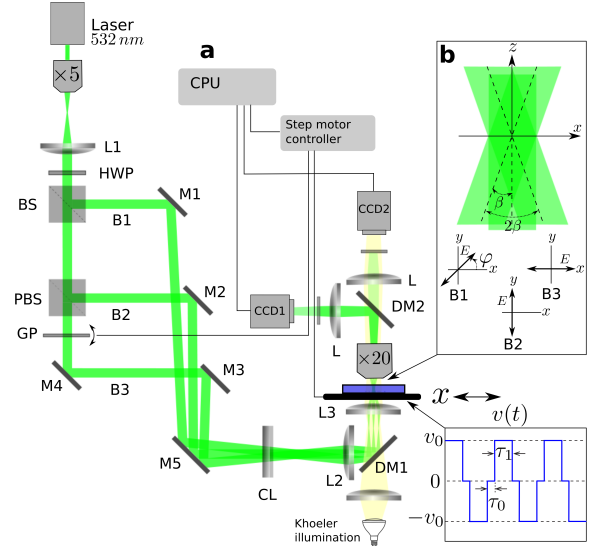}
\vspace{-0.3cm}
\caption{(Color online) (a) Experimental setup: (HWP) half-wave plate, (BS) beam splitter, (PBS) polarizing beam splitter, (M) mirrors, (GP) glass plate, (L) lenses, (CL) cylindrical lens, (DM) dichroic mirrors,  ($\times 5$ and $\times 20$) microscope objectives, ($x$) translation stage, (CCD) cameras. The velocity function driving the translation stage is illustrated on the bottom right. (b) Schematic of the interference by pairs indicating the polarization states of beams 1, 2 and 3.
}\vspace{-0.8cm}\label{fig_setup}
\end{figure}
In order to generate a periodic asymmetric optical lattice, we designed the experimental setup shown schematically in Fig.\ref{fig_setup}. Three beams are interfered by pairs by appropriately setting their respective polarization states in a three-armed Mach-Zehnder interferometer. Two of the beams have orthogonal linear polarization states, while the third one is linearly polarized at an angle $\varphi$ with respect to the horizontal (Fig.~\ref{fig_setup}b), which is set with a half-wave plate (HWP). We generate two superimposed patterns of fringes, one of them with twice the period of the other, determined by the angles $\beta$ and $2\beta$, which can be varied by adjusting the mirrors M1 and M3. The relative intensities of the two patterns can be controlled by the polarization angle $\varphi$ and a relative phase between them can be introduced by tilting a thin glass plate (GP) in one arm of the interferometer. The tilting of GP is done via a motorized actuator. The three beams are directed into a sample cell placed on an XYZ translation stage.

Our samples consist of borosilicate glass microspheres immerse in water with radii in the range of $3.5\,\mu m$ to $7.75\,\mu m$, density of $\rho=2.5\,g/cm^3$ and refractive index $n = 1.56$. For these range of sizes, at room temperature, the thermal fluctuations are negligible \cite{Arzola2009}. We use a laser ($\lambda=532\,nm$) with a fundamental $\text{TEM}_{00}$ emission mode. A cylindrical lens (CL) is used to narrow the resulting pattern in the $y$ direction. The light intensity distribution at the sample plane is described by
\begin{align}\label{eq_Iasim}
I(x, y)=&\frac{2P}{\pi w_x w_y}e^{-2\left(\frac{x^2}{w_x^2}
+\frac{
y^2}{w_y^2}\right)}\left[\sin^2(\varphi)\cos\left(\frac{2\pi}{\Lambda} x\right)
\right.\nonumber\\ 
 &\left.+\cos^2(\varphi)\cos\left(\frac{4\pi}{\Lambda}x+\delta-\pi/2\right)+1\right],
\end{align}
where $P$ denotes the incident optical power at the sample plane. The periods of the two patterns of fringes are $\Lambda$ and $\Lambda/2$, and the width of the Gaussian envelope along the $x$ and $y$ directions are, respectively, $w_x = (745\pm5)\,\mu m$ and $w_y = (19\pm2) \,\mu m$. We are in a regime where $w_x \gg \Lambda$, $w_x \gg w_y$ and $w_x \gg 2R_0$, with $R_0$ the radius of the microspheres. In addition, the dynamics of the particles is observed within the central region of the pattern (about $250\,\mu m$ long). Therefore, we can disregard the effect of the Gaussian envelope along the $x$ direction and consider that we have a 1D optical lattice of period $\Lambda$. The coefficients $\sin^2\varphi$ and $\cos^2\varphi$ are associated with the polarization angle of beam 1 (Fig.~\ref{fig_setup}b), which was set as $\varphi=\pi/4$ in the set of experiments reported here. The parameter $\delta$ represents a phase difference between the two superimposed patterns, and it was chosen so that when $\delta=\pi/2$ they are in phase. The motion of the particles is recorded with a standard video microscopy system (CCD2). An additional camera (CCD1) is used to monitor the light pattern (Fig.~\ref{fig_setup}a).
 
The dynamics of a particle in our system is described by
\begin{equation}\label{eq_model}
 \gamma \dot{x}=-\partial V(x)/\partial x + F_R(t), 
\end{equation}
where we have used the fact that the motion is deterministic and overdamped, with $\gamma$ representing an effective drag coefficient \cite{Arzola2009}. Thus, the time-inversion symmetry is broken. The optical potential is denoted by $V(x)$, and $F_R(t)$ is the rocking force, which acts only along the $x$ direction. The motion of the particle can be considered as 1D along $x$, since it is confined by a strong optical gradient force along the $y$ direction with a single stable equilibrium position, and the weight of the particles is large enough to overcome the scattering optical force along the $z$ direction.

The rocking mechanism is introduced by means of a periodic motion of the translation stage driven with a precision motorized actuator along the direction of the periodicity $x$. The time-periodic force is given by $F_R(t)=\gamma v(t)$, where 

\begin{equation}\label{eq_modgamma}
v(t)=
\begin{cases}
v_0 & \text{if}\ \ 0\leq t< \tau_1,\\
0 & \text{if}\ \ \tau_1\leq t< \tau_1+\tau_0,\\
-v_0 & \text{if} \ \ \tau_1+\tau_0\leq t< T-\tau_0,\\
0 & \text{if} \ \ T-\tau_0\leq t< T.          
\end{cases}
\end{equation}
Here $v_0$ is a constant speed (see bottom right in Fig.~\ref{fig_setup}). The waiting time $\tau_0$ plays a fundamental role in the dynamics, as we shall see below. Importantly, the time-average of $F_R(t)$ over an entire period, $T=2(\tau_0+\tau_1)$, is zero in order to have an unbiased forcing, and thus, the nontrivial ratchet transport.

Recapping, our experimental setup allows the control of the following parameters: the relative intensity of the two periodic patterns (given by $\varphi$), the relative phase between them ($\delta$), the period of the light intensity distribution ($\Lambda$), and the magnitude (via $v_0$) and period ($T$) of the rocking force.  

The  gradient optical force exerted on a particle by a periodic and symmetric pattern of fringes has the same periodicity, but its magnitude depends on the ratio $R_0/\Lambda$ \cite{Arzola2009,ibis,Dholakiacell07,Jonas2008}. In the case of the superposition of two periodic patterns of fringes, the total gradient force acting on a dielectric sphere can be written as 
\begin{align}\label{eq_fuerza}
F(x;\Lambda,R_0)=\frac{P}{c}&\left[A_{\perp}(\Lambda,R_0)\cos\left(\frac{2\pi}{\Lambda} x\right)\right.\nonumber\\
&\left.+A_{\parallel}(\Lambda/2,R_0)\cos\left(\frac{4\pi}{\Lambda}x+\delta\right)\right],
\end{align}
$c$ denoting the light speed in vacuum. The coefficients $A_{\perp}(\Lambda, R_0)$ and $A_{\parallel}(\Lambda/2, R_0)$ determine the optical force for each of the superimposed light lattices with polarizations normal ($\perp$) and parallel ($\parallel$) with respect to the incidence plane. These coefficients vary not only in magnitude but also in sign. We have chosen the definition \eqref{eq_fuerza} in a way that when the value of $A_\perp$ or $A_\parallel$ is positive (negative), the particle is pulled towards the minima (maxima) of the correspondig intensity distribution. We have calculated them using a ray tracing model that we experimentally validated in a previous work for very similar conditions \cite{Arzola2009}. 

From Eq.~\eqref{eq_fuerza}, the optical potential can be expressed as 
\begin{equation}\label{eq_potencial}
V(x;\Lambda,R_0)=-V_0\left[\sin\left(\frac{2\pi}{\Lambda} x\right)+\frac{K}{2}\sin\left(\frac{4\pi}{\Lambda} x+\delta\right)\right],
\end{equation}
where $V_0=P\lvert A_{\perp}\rvert\Lambda/2\pi c$ and $K=A_{\parallel}/\lvert A_{\perp}\rvert$,
for $|A_{\perp}|\neq0$. On the first term on the right hand side of Eq.~\eqref{eq_potencial}, we have ignored a prefactor $\text{sign}(A_{\perp})$ because it leaves invariant the shape of the potential. Importantly, the asymmetry of $V(x)$ is determined by the parameters $K$ and $\delta$. When $K=0.5$ and $\delta=0$, Eq.~\eqref{eq_potencial} describes the typical ratchet potential \cite{Hanggirev09,JLMat2000}. If $K<0$ or $\delta=\pi$, the asymmetry of the potential is inverted, and $\delta=\pm\pi/2$ lead to a symmetric potential. While $\delta$ can be varied at will in our experiment, the value of $K$ depends on the force exerted on the particle by \textit{each} of the two periodic light patterns. When $A_{\perp}\neq0$ and $A_{\parallel}\neq0$, the value of $K$ can be optimized by controlling the relative intensities of the two patterns of fringes via the polarization angle $\varphi$. Notice, however, that the sign of $K$ depends directly on the sign of $A_{\parallel}$, which depends in turn on the radius of the particle for a given period. Therefore, we have found that it is possible to obtain simultaneous opposite motion of particles with different sizes within the same light pattern in our ratchet system. 

\begin{figure}
\includegraphics[angle=0,width=2.7in]{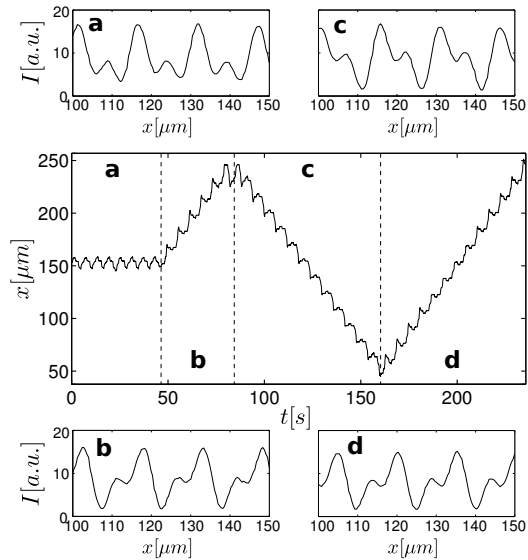}
\vspace{-0.3cm}
\caption{Experimental results for $x$ \textit{vs} $t$ for a particle of radius $R_0= (7.20\pm0.15)\,\mu m$ in an optical lattice of period $\Lambda= (15.3\pm0.1)\,\mu m$. The optical power at the sample is $P=(1.25\pm0.05)\,W$. The parameters of rocking force, Eq.\ref{eq_modgamma}, are: $v_0=(18.8\pm0.5)\,\mu m/s$, $\tau_0=(2.00\pm0.05)\,s$ and $\tau_1=(2.03\pm0.05)\,s$. The different time intervals correspond to: a) $\delta=\pi/2$; b) $\delta=0$; c) $\delta=\pi$ and d) $\delta=0$. Experimental plots of the light intensity distribution for each case are shown on the top and on the bottom.}\vspace{-0.6cm}\label{fig_control}
\end{figure}

In our experimental system, we were able to obtain a directed transport of particles along the light pattern of asymmetric fringes by means of the unbiased external force $F_R(t)=\gamma v(t)$, as defined in Eq.~\eqref{eq_modgamma}. Moreover, we were able to control the direction of motion of the particles in real time by controlling the phase $\delta$. Figure~\ref{fig_control} shows experimental results (video \footnote{See EPAPS Document No. [] for a video file of the experiment described in Fig.~\ref{fig_control}}) for the position of a sphere as a function of time. In the time interval labeled as \textbf{a}, the phase between the two interference patterns is $\delta\approx\pi/2$, giving rise to an approximately symmetric intensity distribution, shown on the top left corner. In the intervals labeled as \textbf{b} and \textbf{d}, the phase was changed to $\delta=0$, giving rise to a positive current (intensity distributions shown in the bottom). Finally, a negative current is observed in the time interval \textbf{c}, for which $\delta\approx\pi$ and hence the asymmetry of the optical lattice is inverted (intensity shown on the top right corner). 

\begin{figure}
\includegraphics[angle=0,width=2.6in]{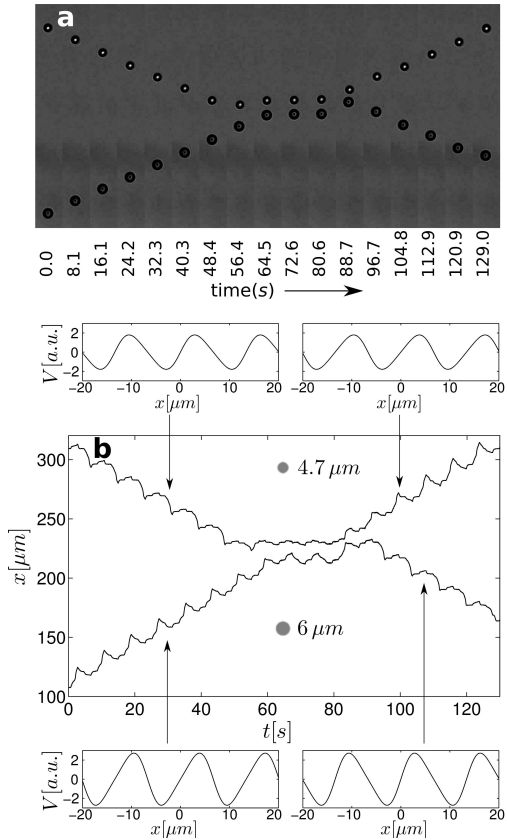}
\vspace{-0.3cm}
\caption{Experimental results for the simultaneous motion of particles of radii: $(6.00\pm0.15)\,\mu m$ and $(4.70\pm0.15)\,\mu m$. The period of the light lattice is $\Lambda=(13.4\pm0.1)\,\mu m$. The optical power at the sample plane was $P=(1.67\pm0.05)\,W$. The parameters of the rocking force are: $v_0=(11.3\pm0.4)\,\mu m/s$, $\tau_0=(1.60\pm0.05)\,s$ and $\tau_1=(1.03\pm0.05)\,s$. (a) Successive frames of the system; the time evolution is indicated at the bottom. (b) $x$ \textit{vs} $t$. The different behaviors observed as the time evolves correspond to different values of the parameter $\delta$. The insets show the calculated optical potential for each particle in the indicated regions.}\vspace{-0.55cm}\label{fig_dosparticulas}
\end{figure}

On the other hand, Fig.~\ref{fig_dosparticulas} shows experimental results (video \footnote{See EPAPS Document No. [] for a video file of the experiment described in Fig.~\ref{fig_dosparticulas}}) for the simultaneous motion of two particles of radii $R_0=(4.70\pm0.15)\,\mu m$ and $R_0=(6.00\pm0.15)\,\mu m$ in an asymmetric light lattice of period $\Lambda=(13.4\pm0.1)\,\mu m$. Successive frames of the two particles indicating the time evolution are shown in Fig.~\ref{fig_dosparticulas}a. Their positions as a function of time are plotted in Fig.~\ref{fig_dosparticulas}b. The two particles are simultaneously moving in opposite directions due to the inverted asymmetry of their corresponding potentials. In the first stage of their paths $\delta\approx0$, and the two particles move towards each other. Then the particles meet at the center of the observation region and they cannot continue their paths. Finally, we change $\delta\approx\pi$ and the particles invert their motion direction, moving apart from each other. The insets indicate the calculated potential experienced by each of the spheres during the initial and final stages of their motion.

\begin{figure}
\includegraphics[angle=0,width=3.4in]{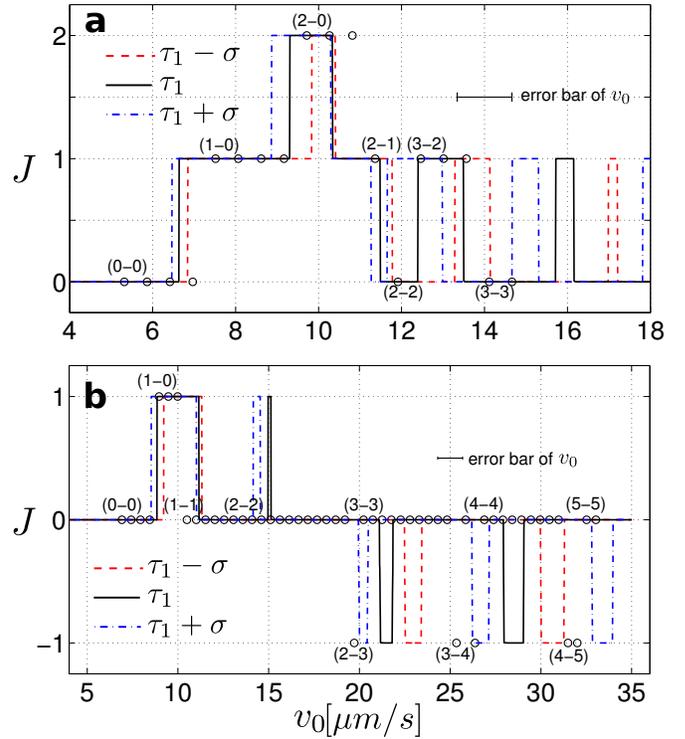}
\vspace{-0.4cm}
\caption{(Color online) Current as a function of $v_0$ for given values of $\tau_1$. The markers represent experimental points and the curves correspond to theoretical calculations for: $\tau_1 - \sigma$ (blue dash-dot), $\tau_1$ (black solid) and $\tau_1 + \sigma$ (red dashed), where (a) $\tau_1=3.43s$, $\sigma = 0.31s$ and (b) $\tau_1=1.70s$, $\sigma_1 = 0.13s$. $\sigma$ denotes the uncertainty in the experimental value of the parameter $\tau_1$. In all cases $\tau_0=(2.00\pm0.05)\,s$.}\vspace{-0.65cm}\label{fig_current}
\end{figure}

The most important parameter characterizing a ratchet system is the particle current, usually defined as the mean particle velocity in stationary conditions: $J=\overline{v(t)}$ for $t\rightarrow\infty$. In our case, the waiting time $\tau_0$ is long enough to allow the particle reaching a stable equilibrium position after each activation semi-cycle $\tau_1$. This means that the particle starts every new cycle with the same initial conditions regarding its relative position in a potential well, and thus we are in a periodic regime. In these circumstances, the current can be expressed as $J=\Delta x/\Lambda$, which has been normalized with respect to $\Lambda/T$. $\Delta x$ represents the net displacement over an entire activation cycle $T$, and it is always given by an integer number of periods. Therefore, $J=n-m$, where $n$ ($m$) is the number of periods the particle is able to move along the direction of lowest (highest) slope in the potential in each semi-cycle. The values of $n$ and $m$ depend on $\tau_1$ and $v_0$ for a given potential. Figure~\ref{fig_current} shows experimental and theoretical results for $J$, including the value of ($n-m$), as a function of $v_0$ for two different values of $\tau_1$. While the structure of discrete jumps exhibited by these plots is a well known phenomenon, the current reversals shown in Fig. 4b, which arise when $m>n$, were  predicted in this regime more than a decade ago \cite{Reimann98} and not previously observed. The role of $\tau_0 \neq 0$ is key for the observation of current reversals, and it may also be key for the eventual observation of chaos in a different regime when $\tau_0 \rightarrow 0$ \cite{Zarlenga09}.

In summary, we have presented the first experimental realization of a deterministic optical rocking ratchet. We obtained a current of particles, whose direction can be controlled by changing the asymmetry of the potential in real time. Our results led us to establish the conditions for observing: 1) simultaneous currents in opposite directions for particles with different sizes in a given light pattern and 2) current reversals for a given particle by varying the magnitude and period of the rocking force. The simplicity and versatility of our system facilitates the comparison with theoretical models, opening the posibility to explore new aspects and solving important questions raised about ratchet dynamics, such as the crucial effect of current reversals.

\begin{acknowledgments}
This project was partially supported by DGAPA-UNAM, grant PAPIIT-IN100110. A. V. Arzola acknowledges support from CONACYT Mexico.
\end{acknowledgments}

\begin{thebibliography}{10}%
\makeatletter
\providecommand \@ifxundefined [1]{%
 \ifx #1\undefined \expandafter \@firstoftwo
 \else \expandafter \@secondoftwo
\fi
}%
\providecommand \@ifnum [1]{%
 \ifnum #1\expandafter \@firstoftwo
 \else \expandafter \@secondoftwo
\fi
}%
\providecommand \enquote [1]{``#1''}%
\providecommand \bibnamefont  [1]{#1}%
\providecommand \bibfnamefont [1]{#1}%
\providecommand \citenamefont [1]{#1}%
\providecommand\href[0]{\@sanitize\@href}%
\providecommand\@href[1]{\endgroup\@@startlink{#1}\endgroup\@@href}%
\providecommand\@@href[1]{#1\@@endlink}%
\providecommand \@sanitize [0]{\begingroup\catcode`\&12\catcode`\#12\relax}%
\@ifxundefined \pdfoutput {\@firstoftwo}{%
 \@ifnum{\z@=\pdfoutput}{\@firstoftwo}{\@secondoftwo}%
}{%
 \providecommand\@@startlink[1]{\leavevmode}%
 \providecommand\@@endlink[0]{}%
}{%
 \providecommand\@@startlink[1]{%
  \leavevmode
  \pdfstartlink
   attr{/Border[0 0 1 ]/H/I/C[0 1 1]}%
   user{/Subtype/Link/A<</Type/Action/S/URI/URI(#1)>>}%
  \relax
 }%
 \providecommand\@@endlink[0]{\pdfendlink}%
}%
\providecommand \url  [0]{\begingroup\@sanitize \@url }%
\providecommand \@url [1]{\endgroup\@href {#1}{\urlprefix}}%
\providecommand \urlprefix [0]{URL }%
\providecommand \Eprint[0]{\href }%
\@ifxundefined \urlstyle {%
  \providecommand \doi [1]{doi:\discretionary{}{}{}#1}%
}{%
  \providecommand \doi [0]{doi:\discretionary{}{}{}\begingroup
  \urlstyle{rm}\Url }%
}%
\providecommand \doibase [0]{http://dx.doi.org/}%
\providecommand \Doi[1]{\href{\doibase#1}}%
\providecommand \bibAnnote [3]{%
  \BibitemShut{#1}%
  \begin{quotation}\noindent
    \textsc{Key:}\ #2\\\textsc{Annotation:}\ #3%
  \end{quotation}%
}%
\providecommand \bibAnnoteFile [2]{%
  \IfFileExists{#2}{\bibAnnote {#1} {#2} {\input{#2}}}{}%
}%
\providecommand \typeout [0]{\immediate \write \m@ne }%
\providecommand \selectlanguage [0]{\@gobble}%
\providecommand \bibinfo [0]{\@secondoftwo}%
\providecommand \bibfield [0]{\@secondoftwo}%
\providecommand \translation [1]{[#1]}%
\providecommand \BibitemOpen[0]{}%
\providecommand \bibitemStop [0]{}%
\providecommand \bibitemNoStop [0]{.\EOS\space}%
\providecommand \EOS [0]{\spacefactor3000\relax}%
\providecommand \BibitemShut [1]{\csname bibitem#1\endcsname}%
%</preamble>
\bibitem{Hanggirev09}%
  \BibitemOpen
  \bibfield{author}{%
  \bibinfo {author} {\bibfnamefont{P.}~\bibnamefont{H{\"a}nggi}}\ and\ \bibinfo
  {author} {\bibfnamefont{F.}~\bibnamefont{Marchesoni}},\ }%
  \bibfield{journal}{%
  \bibinfo {journal} {Rev. Mod. Phys.}\ }%
  \textbf{\bibinfo {volume} {81}},\ \bibinfo {pages} {387} (\bibinfo {year}
  {2009})%
  \bibAnnoteFile{NoStop}{Hanggirev09}%
\bibitem{JLMat2000}%
  \BibitemOpen
  \bibfield{author}{%
  \bibinfo {author} {\bibfnamefont{J.~L.}\ \bibnamefont{Mateos}},\ }%
  \bibfield{journal}{%
  \bibinfo {journal} {Phys. Rev. Lett.}\ }%
  \textbf{\bibinfo {volume} {84}},\ \bibinfo {pages} {258} (\bibinfo {year}
  {2000})%
  \bibAnnoteFile{NoStop}{JLMat2000}%
\bibitem{Faucheux95}%
  \BibitemOpen
  \bibfield{author}{%
  \bibinfo {author} {\bibfnamefont{L.~P.}~\bibnamefont{Faucheux}}, \bibinfo
  {author} {\bibfnamefont{L.~S.}\ \bibnamefont{Bourdieu}}, \bibinfo {author}
  {\bibfnamefont{P.~D.}\ \bibnamefont{Kaplan}},\ and\ \bibinfo {author}
  {\bibfnamefont{A.~J.}\ \bibnamefont{Libchaber}},\ }%
  \bibfield{journal}{%
  \bibinfo {journal} {Phys. Rev. Lett.}\ }%
  \textbf{\bibinfo {volume} {74}},\ \bibinfo {pages} {1504} (\bibinfo {year}
  {1995})%
  \bibAnnoteFile{NoStop}{Faucheux95}%
\bibitem{SHyuk}%
  \BibitemOpen
  \bibfield{author}{%
  \bibinfo {author} {\bibfnamefont{S.-H.}\ \bibnamefont{Lee}}, \bibinfo
  {author} {\bibfnamefont{K.}~\bibnamefont{Ladavac}}, \bibinfo {author}
  {\bibfnamefont{M.}~\bibnamefont{Polin}},\ and\ \bibinfo {author}
  {\bibfnamefont{D.~G.}\ \bibnamefont{Grier}},\ }%
  \bibfield{journal}{%
  \bibinfo {journal} {Phys. Rev. Lett.}\ }%
  \textbf{\bibinfo {volume} {94}},\ \bibinfo {pages} {110601} (\bibinfo {year}
  {2005})%
  \bibAnnoteFile{NoStop}{SHyuk}%
\bibitem{Lopez2008}%
  \BibitemOpen
  \bibfield{author}{%
  \bibinfo {author} {\bibfnamefont{B.~J.}\ \bibnamefont{Lopez}}, \bibinfo
  {author} {\bibfnamefont{N.~J.}\ \bibnamefont{Kuwada}}, \bibinfo {author}
  {\bibfnamefont{E.~M.}\ \bibnamefont{Craig}}, \bibinfo {author}
  {\bibfnamefont{B.~R.}\ \bibnamefont{Long}},\ and\ \bibinfo {author}
  {\bibfnamefont{H.}~\bibnamefont{Linke}},\ }%
  \bibfield{journal}{%
  \bibinfo {journal} {Phys. Rev. Lett.}\ }%
  \textbf{\bibinfo {volume} {101}},\ \bibinfo {pages} {220601} (\bibinfo {year}
  {2008})%
  \bibAnnoteFile{NoStop}{Lopez2008}%
\bibitem{RGommers2006}%
  \BibitemOpen
  \bibfield{author}{%
  \bibinfo {author} {\bibfnamefont{R.}~\bibnamefont{Gommers}}, \bibinfo
  {author} {\bibfnamefont{S.}~\bibnamefont{Denisov}},\ and\ \bibinfo {author}
  {\bibfnamefont{F.}~\bibnamefont{Renzoni}},\ }%
  \bibfield{journal}{%
  \bibinfo {journal} {Phys. Rev. Lett.}\ }%
  \textbf{\bibinfo {volume} {96}},\ \bibinfo {pages} {240604} (\bibinfo {year}
  {2006})%
  \bibAnnoteFile{NoStop}{RGommers2006}%
\bibitem{Salger2009}%
  \BibitemOpen
  \bibfield{author}{%
  \bibinfo {author} {\bibfnamefont{T.}~\bibnamefont{Salger}}, \bibinfo
  {author} {\bibfnamefont{S.}~\bibnamefont{Kling}}, \bibinfo {author}
  {\bibfnamefont{T.}~\bibnamefont{Hecking}}, \bibinfo {author}
  {\bibfnamefont{C.}~\bibnamefont{Geckeler}}, \bibinfo {author}
  {\bibfnamefont{L.}~\bibnamefont{Morales-Molina}},\ and\ \bibinfo {author}
  {\bibfnamefont{M.}~\bibnamefont{Weitz}},\ }%
  \bibfield{journal}{%
  \bibinfo {journal} {Science}\ }%
  \textbf{\bibinfo {volume} {326}},\ \bibinfo {pages} {1241} (\bibinfo {year}
  {2009})%
  \bibAnnoteFile{NoStop}{Renzoni2008}%
\bibitem{Reimann98}%
  \BibitemOpen
  \bibfield{author}{%
  \bibinfo {author} {\bibfnamefont{M.}~\bibnamefont{Schreier}}, \bibinfo
  {author} {\bibfnamefont{P.}\ \bibnamefont{Reimann}}, \bibinfo {author}
  {\bibfnamefont{P.}~\bibnamefont{H{\"a}nggi}},\ and\ \bibinfo {author}
  {\bibfnamefont{E.}~\bibnamefont{Pollak}},\ }%
  \bibfield{journal}{%
  \bibinfo {journal} {Europhys. Lett.}\ }%
  \textbf{\bibinfo {volume} {44}},\ \bibinfo {pages} {416} (\bibinfo {year}
  {1998})%
  \bibAnnoteFile{NoStop}{Reimann98}%
\bibitem{Zarlenga09}%
  \BibitemOpen
  \bibfield{author}{%
  \bibinfo {author} {\bibfnamefont{D.~G.}~\bibnamefont{Zarlenga}}, \bibinfo
  {author} {\bibfnamefont{H.~A.}\ \bibnamefont{Larrondo}}, \bibinfo {author}
  {\bibfnamefont{C.~M.}~\bibnamefont{Arizmendi}},\ and\ \bibinfo {author}
  {\bibfnamefont{F.}~\bibnamefont{Family}},\ }%
  \bibfield{journal}{%
  \bibinfo {journal} {Phys. Rev. E}\ }%
  \textbf{\bibinfo {volume} {80}},\ \bibinfo {pages} {011127} (\bibinfo {year}
  {2009})%
\bibitem{aldana2006}%
  \BibitemOpen
  \bibfield{author}{%
  \bibinfo {author} {\bibfnamefont{R.}~\bibnamefont{Salgado-Garc{\'i}a}}, 
  \bibinfo {author} {\bibfnamefont{M.}~\bibnamefont{Aldana}}\ and\
  \bibinfo {author} {\bibfnamefont{G.}~\bibnamefont{Martinez-Mekler}},\ }%
  \bibfield{journal}{%
  \bibinfo {journal} {Phys. Rev. Lett.}\ }%
  \textbf{\bibinfo {volume} {96}},\ \bibinfo {pages} {134101} (\bibinfo {year}
  {2006})%
  \bibAnnoteFile{NoStop}{aldana2006}%
  \bibAnnoteFile{NoStop}{Zarlenga09}%
\bibitem{Arzola2009}%
  \BibitemOpen
  \bibfield{author}{%
  \bibinfo {author} {\bibfnamefont{A.~V.}\ \bibnamefont{Arzola}}, \bibinfo
  {author} {\bibfnamefont{K.}~\bibnamefont{Volke-Sep{\'u}lveda}},\ and\
  \bibinfo {author} {\bibfnamefont{J.~L.}\ \bibnamefont{Mateos}},\ }%
  \bibfield{journal}{%
  \bibinfo {journal} {Opt. Express}\ }%
  \textbf{\bibinfo {volume} {17}},\ \bibinfo {pages} {3429} (\bibinfo {year}
  {2009})%
  \bibAnnoteFile{NoStop}{Arzola2009}%
\bibitem{ibis}%
  \BibitemOpen
  \bibfield{author}{%
  \bibinfo {author} {\bibfnamefont{I.}~\bibnamefont{Ric{\'a}rdez-Vargas}},
  \bibinfo {author} {\bibfnamefont{P.}~\bibnamefont{Rodr{\'i}guez-Montero}},
  \bibinfo {author} {\bibfnamefont{R.}~\bibnamefont{Ramos-Garc{\'i}a}},\ and\
  \bibinfo {author} {\bibfnamefont{K.}~\bibnamefont{Volke-Sep{\'u}lveda}},\ }%
  \bibfield{journal}{%
  \bibinfo {journal} {Appl. Phys. Lett.}\ }%
  \textbf{\bibinfo {volume} {88}},\ \bibinfo {pages} {121116} (\bibinfo {year}
  {2006})%
  \bibAnnoteFile{NoStop}{ibis}%
\bibitem{Dholakiacell07}%
  \BibitemOpen
  \bibfield{author}{%
  \bibinfo {author} {\bibfnamefont{K.}~\bibnamefont{Dholakia}}, \bibinfo
  {author} {\bibfnamefont{M.~P.}\ \bibnamefont{MacDonald}}, \bibinfo {author}
  {\bibfnamefont{P.}~\bibnamefont{Zem{\'a}nek}},\ and\ \bibinfo {author}
  {\bibfnamefont{T.}~\bibnamefont{\v{C}i\v{z}m{\'a}r}},\ }%
  \bibfield{journal}{%
  \bibinfo {journal} {Methods Cell Biol}\ }%
  \textbf{\bibinfo {volume} {82}},\ \bibinfo {pages} {467} (\bibinfo {year}
  {2007})%
  \bibAnnoteFile{NoStop}{Dholakiacell07}%
\bibitem{Jonas2008}%
  \BibitemOpen
  \bibfield{author}{%
  \bibinfo {author} {\bibfnamefont{A.}~\bibnamefont{Jon{\'a}\v{s}}}\ and\
  \bibinfo {author} {\bibfnamefont{P.}~\bibnamefont{Zem{\'a}nek}},\ }%
  \bibfield{journal}{%
  \bibinfo {journal} {Electrophoresis}\ }%
  \textbf{\bibinfo {volume} {29}},\ \bibinfo {pages} {4813} (\bibinfo {year}
  {2008})%
  \bibAnnoteFile{NoStop}{Jonas2008}%
\bibitem{Note1}%
  \BibitemOpen
  \bibinfo {note} {See EPAPS Document No. [Ref15.avi] for a video of the experiment
  described in Fig.~\ref {fig_control} (speeded up by a factor of 5). The top and bottom halves were captured 
with cameras CCD1 and CCD2, respectively. The experimental intensity profile is also 
shown in the top video; the changes in the light lattice are produced when 
$\delta$ varies}%
  \bibAnnoteFile{NoStop}{Note1}%
\bibitem{Note2}%
  \BibitemOpen
  \bibinfo {note} {See EPAPS Document No. [Ref16.avi] for a video of the experiment
  shown in Fig.~\ref {fig_dosparticulas} (speeded up by a factor of 5)}%
  \bibAnnoteFile{NoStop}{Note2}%
\end{thebibliography}
\end{document}